\documentclass[runningheads]{svmult}
\usepackage{amsmath}     
\usepackage{amssymb}    
\usepackage{epsfig}     
\def \BE {\begin{equation}}
\def \EE {\end{equation}}
\def \BEA {\begin{eqnarray}}
\def \EEA {\end{eqnarray}}

\def \k {\vec k}
\def \q {\vec q}
\def \x {\vec x}
\def \kx {{\vec k} \cdot {\vec x}}

\newcommand{\EM}[1]{\emph{#1}}%                  % italic text font
\newcommand{\PA}[1]{\partial_{#1}}%              % partial deriv
\newcommand{\Dt}{D_{t}}%                         % material deriv 

\begin{document}
\baselineskip 20pt
\title*{Wave turbulence in Bose-Einstein condensates}
\toctitle{Wave turbulence in Bose-Einstein condensates}
\titlerunning{Wave turbulence in Bose-Einstein condensates}
\author{Yuri Lvov\inst{1} \and Sergey Nazarenko\inst{2} \and Robert West\inst{2}}
\authorrunning{Lvov, Nazarenko, West}
\institute{Department of Mathematical Sciences, Rensselaer Institute for 
           Mathematical Sciences, New York, 12180-3590, USA. \and 
           Mathematics Institute, University of Warwick, Coventry, CV4 7AL, UK.}

\maketitle

%%%%%%%%%%%%%%%%%%%%%%%%%%%%%%%%%%%%%%%%%%%%%%%%%%%%%%%%%%
%%%%%%%%%%%%%%%%%%%%% ABSTRACT %%%%%%%%%%%%%%%%%%%%%%%%%%%
%%%%%%%%%%%%%%%%%%%%%%%%%%%%%%%%%%%%%%%%%%%%%%%%%%%%%%%%%%

\begin{centerline}{\cal{ To Appear in Physica D}}\end{centerline}

\begin{abstract}
The kinetics of nonequilibrium Bose-Einstein condensates are
considered within the framework of the Gross-Pitaevskii equation.  A
systematic derivation is given for weak small-scale perturbations of a
steady confined condensate state. This approach combines a wavepacket
WKB description with the weak turbulence theory. The WKB theory
derived in this paper describes the effect of the condensate on the
short-wave excitations which appears to be different from a simple
renormalization of the confining potential suggested in previous
literature.
\end{abstract}

%%%%%%%%%%%%%%%%%%%%%%%%%%%%%%%%%%%%%%%%%%%%%%%%%%%%%%%%%%
%%%%%%%%%%%%%%%%%% MAIN DOCUMENT %%%%%%%%%%%%%%%%%%%%%%%%%
%%%%%%%%%%%%%%%%%%%%%%%%%%%%%%%%%%%%%%%%%%%%%%%%%%%%%%%%%%

%%%%%%%%%%%%%%%%%%%%%%%%%%%%%%%%%%
\section{Introduction}\label{sec1}
%%%%%%%%%%%%%%%%%%%%%%%%%%%%%%%%%%

Bose-Einstein condensate (BEC) was first observed in 1995 in atomic
vapors of $^{87}$Rb~\cite{Anderson}, $^7$Li~\cite{Bradley} and
$^{23}$Na~\cite{Davis}. Typically, the gas of atoms is confined by a
magnetic trap~\cite{Anderson}, and cooled by laser and evaporative
means.  Although the basic theory for the condensation was known from
the classical works of Bose~\cite{Bose} and Einstein~\cite{Einstein},
the experiments on BEC stimulated new theoretical work in the field
(an excellent review of this material is given in~\cite{Pitaevsky}).

\begin{figure}[b]
  \begin{center}
  \includegraphics[width=.6\textwidth]{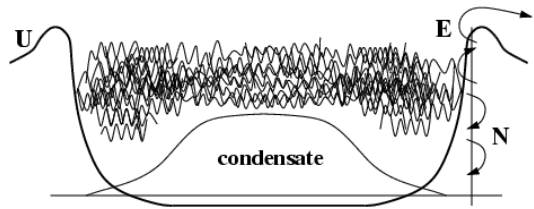}
  \end{center}
  \caption[]{Turbulent cascades of energy $E$ and particle number
  $N$.}
  \label{fig1}
\end{figure}

A lot of theoretical results about condensate dynamics are based on
the assumption that the condensate band can be characterized by some
temperature $T$ and chemical potential $\mu$, the quantities which are
clearly defined only for gases in thermodynamic equilibrium.  Often,
however, the condensation is so rapid that the gas is in a very
nonequilibrium state and hence, one requires the use of a kinetic
rather than a thermodynamic theory \cite{Gardiner,Gardiner2,SV}.  An
approach using the quantum kinetic equation was developed by Gardiner
et al \cite{Gardiner,Gardiner2} who used some phenomenological
assumptions about the scattering amplitudes. Phenomenology is
unavoidable in the general case due to an extreme dynamical complexity
of quantum gases the atoms in which interact among themselves and
exhibit wave-particle dualism. Most phenomenological assumptions are
intuitive or arise from a physical analogy and are hard to validate
(or to prove wrong) theoretically. In particular, it was proposed that
the ground BEC states act onto the higher levels via an effective
potential.  In the present paper we are going to examine this
assumption in a special case of large occupation numbers, i.e. when
the system is more like a collection of interacting waves rather than
particles and which allows a systematic theoretical treatment.
In what follows we show {\it systematically} that such an
assumption is not true for such systems. For dilute
gases, with a large number of atoms at low temperatures, one obtains
the Gross-Pitaevskii (GP) equation for the condensate order
parameter~\cite{Gross,Pitaevsky1961}:
\begin{equation}\label{eq1}
    i \PA{t}\psi +\triangle \psi - |\psi|^{2}\psi -U\psi= 0,
\end{equation}
where the potential $U$ is a given function of coordinate, see for
example figure \ref{fig1}. 
%In the limit $U=0$, the GP
%equation becomes the defocusing Nonlinear Schr\"odinger (NLS) equation.
We emphasize that the area of validity of GP equation is restricted to
a narrow class of the low-temperature BEC growth experiments and the
latest stages in other BEC experiments. However, we will study the GP
equation because it provides an important limiting case for which one
can rigorously test the phenomenological assumptions made for more
general systems. We would like to abandon the approach where the system is
artificially divided into a $T=0$ condensate state and a thermal ``cloud'' because
this ``cloud'' in reality is far from the thermodynamic equilibrium and
we believe that this fact affects the BEC dynaimcs in an essential way.
As in many other non-equilibrium and turbulent systems, fluxes of the conserved
quantities through the phase space are more relevant for the theory here than 
the temperature and the chemical potential. Performance of a thermodynamic theory
here would be as poor as a description of waterfalls by a theory developed for lakes.\footnote{
This comparison was suggested by Vladimir Zakharov to illustrate irrelevance of the
thermodynamic approach to the turbulence of dispersive waves.}
Again, the GP equation is used in our work for both the ground and the excited states
which limits our analysis only to the low temperature and high occupation number
situations. 

In fact the idea of using GP equation for describing BEC kinetics is not new and it 
goes back to work
of Kagan et al \cite{SV}, who used a kinetic equation for waves
systematically derived from the GP equation ignoring the trapping
potential and assuming turbulence to be spatially homogeneous
\cite{ZMR85}.  A similar method has been used to investigate optical
turbulence \cite{DNPZ92}.  Classical weak turbulence theory yields a
closed kinetic equation for the long time behavior of the energy
spectrum without having to make unjustifiable assumptions about the
statistics of the processes
\cite{ZLF,Newell,Ben,NazarenkoNewell,Newell68,LBN}.  Second, the kinetic
equation admits classes of {\it exact} equilibrium solutions
\cite{ZLF,Z68a,Z68b}. These can be identified as pure Kolmogorov
spectra \cite{ZMR85,DNPZ92,ZLF}, namely equilibria for which
there is a constant spectral flux of one of the invariants, the
energy,
   $$E = \int [|\nabla \psi|^2 + \frac{1}{2} |\psi|^4] \, d {\bf x},$$ 
and the ``number of particles'',
   $$N =\int |\psi|^2 \, d {\bf x}.$$ 
A very important property of the particle cascade is that it transfers
the particles to the small $k$ values (inverse cascade).  This
transfer will lead to an accumulation at small $k$'s which is
precisely the mechanism of the BE condensation, see figure 1. The
energy cascade is toward high values of $k$ which eventually will lead
to ``spilling'' over the potential barrier corresponding to an
evaporative cooling, see figure 1.  After the formation of strong
condensate one can no longer use weak turbulence theory, as the weak
turbulence theory assumes small amplitudes. However, one can
reformulate the theory using a linearization around the condensate,
(as oppose to linearization around the $0$ state), as in
\cite{DNPZ92}. Consequently this changes the dominant system
interactions from 4-wave to 3-wave processes.

Kolmogorov-type energy distributions over the levels (scales) are
dramatically different from any thermodynamic equilibrium
distributions. Thus, the condensation and the cooling rates will also
be significantly different from those obtained from theories based on
the assumptions of a thermodynamic equilibrium and the existence of a
Boltzmann distribution. As an example, a finite-time condensation was
predicted by Kagan, Svistunov and Shlyapnikov~\cite{SV}, whose work
was based on the theory of weak homogeneous turbulence.

However, application of the theory of homogeneous turbulence to the GP
equation has its limitations.  Indeed, when the external potential is
not ignored in the GP equation, the turbulence is trapped and is,
therefore, intrinsically inhomogeneous (e.g. a turbulent
spot). Additional inhomogeneity of the turbulence arises because of
the condensate, which in the GP equation case is itself coordinate
dependent.  This means, in particular, that the theory of homogeneous
turbulence cannot describe the ground state effect onto the confining
properties of the gas and thereby test the effective potential
approach.  The present paper is aimed at removing this pitfall via
deriving an inhomogeneous weak turbulence theory.

The effects of the coordinate dependent potential and condensate can
most easily be understood using a wavepacket (WKB) formalism that is
applicable if the wavepacket wavelength $l$ is much shorter than the
characteristic width of the potential well $L$,
\begin{equation*}
  \varepsilon = \frac{l}{L} \ll 1.
\end{equation*}
The coordinate dependent potential and the condensate distort the
wavepackets so that their wavenumbers change.  This has a dramatic
effect on nonlinear resonant wave interactions because now waves can
only be in resonance for a finite time.  The goal of our paper is to
use the ideas developed for the GP equation without the trapping
potential and to combine them with the WKB formalism in order to
derive a weak turbulence theory for a large set of random waves
described by the GP equation.

Note that idea to combine the kinetic equation with WKB to describe
weakly nonlinear dynamics of wave (or quantum) excitations is quite
old and can be traced back to Khalatnikov's theory of Bose gas (1952)
and Landau's theory of the Fermi fluids (1956), see e.g. in
\cite{landafshits10}. It has also been widely used to describe
kinetics of waves in plasmas, e.g. \cite{bbk,tsit1,tsit2,zmrub}. For
plasmas, such a formalism was usually derived from the first
principles. However, only phenomenological models based on an
experimentally measured dispersion curves have been proposed so far
for the superfluid kinetics.  In this paper, we offer for the first
time a consistent derivation starting from the GP equation which
allows us to correct the existing BEC phenomenology at least for the
special cases when the GP equation is applicable.

Technically, the most nontrivial new element of our theory appears
through the linear dynamics (WKB) whereas modifications of the
nonlinear part (the collision integral) are fairly straightforward.
Thus, we start with a detailed consideration of the linear dynamics in
section \ref{sec2}. Previously, linear excitations to the ground state
were considered by Fetter \cite{fetter} who used a test function approach 
to derive an approximate dispersion relation for these excitations.
Fetter pointed out an uncertainty of the boundary conditions to be used
at the ground state reflection surface. The WKB theory for BEC which is 
for the first time developed in the present paper allows an asymptotically
rigorous approach which, among other things, allows to clarify the role
of the ground state reflection surface.
Indeed, as we will see in section 3,  the  WKB theory
is essentially different in the case when the condensate ground state
is weak and can be neglected from the case of strongly nonlinear
ground state. No suitable WKB description exists for the intermediate
case in which the linear and the nonlinear effects are of the same
order. However, in the Thomas-Fermi regime the layer of the
intermediate condensate amplitudes is extremely narrow due to the
exponential decay of the amplitude beyond the ground state reflection
surface.  This allowed us to combine the two WKB descriptions into one
by formally re-writing the equations in such a way that they are
correct in the limits of both weak and strong condensate. These
equations will be wrong in the thin layer of intermediate condensate
amplitudes, but this will not have any effect on the overall dynamics
of wavepackets because they pass this layer too quickly to be affected
by it. 

In section 4 for the first time we present a  Hamiltonian formulation
of the WKB equations and derive a cannonical Hamiltonian the form
of which is general for all WKB systems and not only BEC. 
The Hamiltonian formulation
is needed to prepare the scene for the weak turbulence theory. In
section 5 we apply weak turbulence theory to write a closed kinetic
equation for wave action. This kinetic equation has a coordinate
dependence of the frequency delta functions. Notice that coordinate
dependence of the wave frequency has a profound effect on the
nonlinear dynamics. The resonant wave interactions can now take place
only over a limited range of wave trajectories which makes such
interactions similar to the collision of discrete particles.

%%%%%%%%%%%%%%%%%%%%%%%%%%%%%%%%%%%%%%%%%%%%%%%%%%%%%%%%%%
\section{Linear dynamics of the GP equation \label{sec2}}
%%%%%%%%%%%%%%%%%%%%%%%%%%%%%%%%%%%%%%%%%%%%%%%%%%%%%%%%%%

We will now develop a WKB theory for small-scale wave-packets,
described by a linearized GP equation, with and without the presence
of a background condensate. As is traditional with any WKB-type method
we assume the existence of a scale separation $\varepsilon \ll 1$, as
explained in section \ref{sec1}. In this analysis we will take $l\sim
1$ so that any spatial derivatives of a given large-scale quantity
(e.g. the potential $U$ or the condensate) are of order
$\varepsilon$. The transition to WKB phase-space is achieved through
the application of the Gabor transform \cite{nkd},
\begin{equation}\label{Gabor}
   \hat{g}(\vec{x},\vec{k},t) = \int
    f(\varepsilon^{*}|\vec{x} - \vec{x}_{0}|) \,
    \E^{i\vec{k}\cdot(\vec{x} - \vec{x}_{0})} \, g(\vec{x}_{0},t)\,d\vec{x}_{0},
\end{equation}
where $f$ is an arbitrary function fastly decaying at infinity.
For our purposes it will be sufficient to consider a Gaussian of the form
   $$f(\vec{x}) = \frac{1}{(2\pi)^{d}}\,e^{- x^2},$$
where $d$ is the number of space dimensions. The parameter
$\varepsilon^{*}$ is small and such that $\varepsilon \ll
\varepsilon^{*} \ll 1$. Hence, our kernel $f$ varies at the
intermediate-scale. A Gabor transform can therefore be thought of as a
localized Fourier transform, and in the limit $\epsilon^*\to 0$
becomes an exact Fourier transform.  Physically, one can view a Gabor
transform as a wavepacket distribution function over positions
$\vec{x}$ and wavevectors $\vec{k}$.

%---------------------------------------------%
\subsection{Linear theory without a condensate}
%---------------------------------------------%
Linearizing the GP equation, to investigate the behavior of
wavepackets $\psi$ without the presence of a condensate, we obtain the
usual linear Schr\"odinger equation:
\begin{equation}\label{LGPENC}
   i \PA{t}\psi + \triangle \psi -U\psi = 0,
\end{equation}
where $U$ is a slowly varying potential. Let us apply the Gabor
transformation to (\ref{LGPENC}). Note that the Gabor transformation
commutes with the Laplacian, so that $\widehat{\Delta \Psi}=\Delta
\hat{\Psi}$.  Also note that 
  $$\widehat{U\Psi}\simeq U \hat{\Psi} + i  (\nabla_x U)\nabla_k\hat{\Psi},$$ 
where we have neglected the quadratic and higher order terms in
$\epsilon$ because $\Psi$ changes on a much shorter scale than the
large scale function $U$. Combining the Gabor transformed equation
with its complex conjugate we find the following WKB transport
equation,
\begin{equation}\label{phi}
   \Dt |\hat{\psi}|^{2} = 0,
\end{equation}
where 
  $$\Dt \equiv \PA{t} + \dot{\vec{x}}\cdot\nabla +\dot{\vec{k}}\cdot\PA{k},$$
represents the total time derivative along the wavepacket trajectories
in phase-space.  The ray equations are used to describe wavepacket
trajectories in $(\vec{k},\vec{x})$ phase-space,
\begin{align}\label{rays}
   &\dot{\vec{x}} = \PA{k}\omega,
   &\dot{\vec{k}} = -\nabla\omega.
\end{align}
The frequency $\omega$, in this case, is given by $\omega = k^{2}+U$,
(again we use the notation $k=|\vec{k}|$).  Equations (\ref{phi}) and
(\ref{rays}) are nothing more than the famous \EM{Ehrenfest} theorem
from quantum mechanics. According to (\ref{rays}), the wavepackets
will get reflected by the potential at points $r_{R}$ where $U(r_{R})
= k^{2}_{max}$. We will now move on to consider linear wavepackets in
the presence of a background condensate.

%---------------------------------------------------------%
\subsection{Wavepacket dynamics on a condensate background}
%---------------------------------------------------------%
One of the common assumptions in the BEC theory is that the presence
of a condensate acts on the higher levels by just modifying the
confining potential $U$, see for example~\cite{GLBDZ98}.  If this was
the case, the linear dynamics would still be described by the
\EM{Ehrenfest} theorem with some new effective potential. We will show
below that this is not the case.

Let us define the condensate $\psi_{0}$ as a nonlinear coordinate
dependent solution of equation (\ref{eq1}), with a lengthscale of the
order of the ground state size (although it does not need to be
exactly the same as the ground state).  In what follows, we will use
Madelung's amplitude-phase representation for $\psi_{0}$, namely
\begin{equation}
   \psi_{0} = \sqrt{\rho({\bf r})} \, e^{ i \theta},
   \label{condensaterepresentation}
\end{equation}
where ${\bf v} = 2 \nabla\theta$ is the macroscopic speed of the
condensate.  It is well known that in this representation $\rho$ obeys
a continuity equation,
\begin{equation}
   \rho_t+ \hbox{div}({\rho {\bf v}}) = 0.
   \label{continuity}
\end{equation}
For future reference, one should note that the second term in this
expression is $O(\epsilon^2)$. Thus, $\rho_t$ is $O(\epsilon^2)$ too
and it must be neglected in the WKB theory which takes into account
only linear in $\epsilon$ terms.  We start by considering a small
perturbation $\phi \ll 1$, such that
\begin{equation}
   \psi = \psi_{0}(1+\phi).
   \label{YuriSubstitution}
\end{equation}
Substituting (\ref{YuriSubstitution}) into (\ref{eq1}) we find
\begin{equation}\label{eq3}
    i  \PA{t}\phi + \triangle\phi 
    + 2\frac{\nabla\psi_{0}}{\psi_{0}}\cdot\nabla\phi
    -  \varrho\Big(\phi+\phi^{*} +2 |\phi|^2 +\phi^2 +|\phi|^2\phi\Big)=0.  
\end{equation}
where $\varrho=\varrho(\vec{x})=|\psi_{0}|^{2}$ is a slowly varying
condensate density.

In a similar manner to the previous subsection, the rest of this
derivation consists of Gabor transforming (\ref{eq3}), combining the
result with its complex conjugate and finding a suitable waveaction
variable such that the transport equation represents a conservation
equation along the rays. Such a derivation is given in Appendix A.  It
yields to the following expression for the waveaction,
\begin{equation} 
   n({\bf k},x,t) = \frac{1}{2} \frac{\omega\rho}{k^{2}} 
                    \left| \widehat{ \Re \phi} 
                  - \frac{ i k^{2} }{\omega} \widehat{ \Im \phi} 
                    \right|^{2},
   \label{wacconden}
\end{equation}
where $\Re$ and $\Im$ mean the real and imaginary parts respectively.
As usual, the transport equation takes the form of a conservation
equation for waveaction along the rays,
\begin{equation}
   \Dt n(\vec{x},\vec{k},t) = 0,
   \label{WCondensateWKB}
\end{equation}
where
\begin{equation}
   \Dt \equiv \PA{t} 
 + \dot{\vec{x}}\cdot\nabla 
 + \dot{\vec{k}}\cdot\PA{k},  
\end{equation}
is the time derivative along trajectories 
\begin{equation}
   \dot{\vec{x}} = \PA{k} \omega, \hspace{1cm}
   \dot{\vec{k}} = -\nabla \omega.
\end{equation}
The frequency is given by the following expression,
\begin{equation}
   \omega = k \sqrt{k^{2}+2\varrho}.
   \label{OmCondensate}
\end{equation}
One can immediately recognize in (\ref{OmCondensate}) the Bogolubov's
formula \cite{Bogolubov1947} which was derived before for systems with
a coordinate independent condensate and without a trapping
potential. It is remarkable that presence of the potential $U$ does
not affect the frequency so that expression (\ref{OmCondensate})
remains the same.  Obviously, the dynamics in this case cannot be
reduced to the \EM{Ehrenfest} theorem with any shape of potential
$U$. Therefore, an approach that models a condensate's effect by
introducing a renormalized potential would be misleading in this case.

%%%%%%%%%%%%%%%%%%%%%%%%%%%%%%%%%%%%%%%%%%%
\section{Applicability of WKB descriptions}
%%%%%%%%%%%%%%%%%%%%%%%%%%%%%%%%%%%%%%%%%%%
%
\begin{figure}[b]\label{fig2}
  \begin{center}
  \includegraphics[width=.6\textwidth]{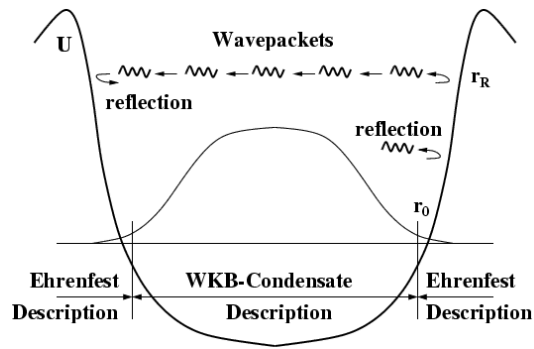}
  \end{center}
  \caption{Regions of Applicability of WKB Descriptions.}  
\end{figure}    
In this section we will investigate the applicability of the above
theory. Let us consider a condensate which is a solution of the
eigenvalue problem $\PA{t}\psi_{0} = -i \Omega \psi_{0}$. Therefore,
the GP equation (\ref{eq1}) becomes
\begin{equation}
   \Omega \psi_{0}+\triangle\psi_{0}-\varrho\psi_{0} - U\psi_{0} = 0.
   \label{evp}
\end{equation} 
%

%-------------------------------%
\subsection{Weak condensate case}
%-------------------------------%
Firstly, let us consider the case of a weak condensate so that the
effect of the nonlinear term is small in comparison to the linear
ones, $|\varrho\psi_{0}|\lesssim|\triangle\psi_{0}|$.  Since $\Omega$
is a constant we observe that the Laplacian term acts to balance the
external potential term (like in the linear Schr\"odinger equation)
and the nonlinear term can be, at most, as big as the linear ones
  $$\Omega \sim \frac{1}{r_{0}^{2}} \sim U(r_{0}) \gtrsim \varrho,$$ 
where $r_{0}$ is the characteristic size of the condensate (it is defined 
as the condensate \lq\lq reflection'' point via the condition $\Omega 
=U(r_{0})$, see below).

Now for a WKB description to be valid we require $kr_{0}\gg 1$,
i.e. we require the characteristic length-scale of our wavepackets to
be a lot smaller than that of the large-scales. Using this fact we
find
  $$k^{2} \gg \frac{1}{r_{0}^{2}} \sim U(r_{0}) \gtrsim \varrho.$$
Therefore, the condensate correction to the frequency, given by
(\ref{OmCondensate}), is small. In other words the wavepacket
does not \lq \lq feel'' the condensate. Indeed, from $k_{max}^{2} =
U(r_{R})$ we have $U(r_{R}) \gg U(r_{0})$ and this implies that $r_{R}
\gg r_{0}$ (where $r_{R}$ is the wavepacket reflection point, see
figure 2). Thus, the condensate in this case occupies a tiny space at
the bottom of the potential well and hence does not affect a
wavepacket's motion. Therefore, a wavepacket moves as a \lq\lq
classical'' particle described by the \EM{Ehrenfest} equations
(\ref{phi}) and (\ref{rays}).  In fact, in this case it would be
incorrect to try to describe the small condensate corrections via our
WKB approach because these corrections are of order $\varrho \sim
\varepsilon^{2}$ (the $\varepsilon^{2}$ terms being ignored in a WKB
description).

%---------------------------------%
\subsection{Strong condensate case}
%---------------------------------%
Now we will consider a strong condensate such that
\begin{equation}\label{cond2}
   \Omega \cong U 
 + \varrho \gg \frac{|\triangle\psi_{0}|}{|\psi_{0}|},
\end{equation}
i.e. the $r$ dependence of the potential $U$ is now balanced by the
nonlinearity.  This is usually referred to as the \EM{Thomas-Fermi}
limit~\cite{Pitaevsky}. Wavepackets now \lq\lq feel'' the presence of
a strong condensate if $\varrho \sim k^{2}$.  We see that the WKB
approach is applicable because
 $$k^{2}\sim \varrho \gg \frac{1}{r_{0}^{2}} \sim 
           \frac{|\triangle\psi_{0}|}{|\psi_{0}|}.$$
According to the ray equations $\omega$ is a constant along a
wavepacket's trajectory, so we can find the packet's wavenumber from
$k^{2} = \sqrt{\varrho^{2}+\omega^{2}}-\varrho$.  One can see that
$k^2$ remains positive for any value of $\varrho$ which means that the
presence of the condensate does not lead to any new wavepacket
reflection points (i.e. when $k$ takes a value of zero). Thus,
turbulence is allowed to penetrate into the center of the potential
well.  However, the group velocity increases when the condensate
becomes stronger, $\partial_k \omega \sim \sqrt \rho$. This means that
the density of wavepackets decreases toward the center of
well. Therefore, the condensate tends to push the turbulence away from
the center, toward the edges of the potential trap.

To summarize, in the presence of a strong condensate we have two
regions of applicability for our WKB descriptions, see figure 2.
Wavepackets at a position $r<r_{0}$, in the central region of the
potential well will evolve according to the WKB-condensate description
(\ref{wacconden}) - (\ref{OmCondensate}).  The Laplacian term only
becomes important for $r>r_{0}$ where $\varrho$ is exponentially
small. In this case the Ehrenfest description is appropriate. It will
be shown in the next section that these two WKB descriptions can be
combined into a single set of formulae.

%----------------------------------%
\subsection{Unified WKB description}
%----------------------------------%
It is interesting  that taking the limit of zero condensate amplitude
in the waveaction (\ref{wacconden}) results in the waveaction
$\frac{1}{2}|\hat\Psi|^2$ of the Ehrenfest equation (\ref{phi})
which corresponds to the regime without condensate,
  $$ \lim\limits_{\rho\to 0}n({\bf k},x,t)\to 
     \frac{1}{2}\rho \left| \widehat{\phi}\right|^{2}
   = \frac{1}{2}|\hat\Psi|^2.$$
On the other hand, $ \lim\limits_{\rho\to 0} \omega \to k^2 $ which is
different from the Ehrenfest expression $ \omega = k^2 +U $.  Thus,
one cannot recover the non-condensate (Ehrenfest) description by just
taking the limit of zero condensate amplitude in (\ref{wacconden}),
(\ref{WCondensateWKB}) and  (\ref{OmCondensate}).
However, one can easily write a unified WKB description which will be
valid with or without condensate by simply adding $U+\rho$ to the
frequency (\ref{OmCondensate}). Indeed, for strong condensate
$U+\rho$=const and, therefore, it does not alter the ray equations
(which contain only derivatives of $\omega$).  On the other hand, such
an addition allows us to obtain the correct expression
 $$ \omega = k^2 +U,$$
in the limit $\rho\to 0$.  Summarizing, we write the following
equations of the linear WKB theory which are valid with or without the
presence of a condensate,
\begin{equation}
   \Dt n(\vec{x},\vec{k},t) = 0,
   \label{WCondensateWKBu}
\end{equation}
where
\begin{equation} n({\bf k},x,t) = \frac{1}{2}
   \frac{\omega\rho}{k^{2}} \left| \widehat{ \Re \phi} -
   \frac{ i k^{2} }{\omega} \widehat{ \Im \phi} \right|^{2},
   \label{waccondenu}
\end{equation} 
is the waveaction and 
\begin{equation}
   \Dt \equiv \PA{t} + \dot{\vec{x}}\cdot\nabla 
                     + \dot{\vec{k}}\cdot\PA{k},
   \label{fdar}  
\end{equation}
is the full time derivative along trajectories and
\begin{equation}
   \dot{\vec{x}} = \PA{k} \omega, \hspace{1cm}
   \dot{\vec{k}} = -\nabla \omega,
\end{equation}
are the ray equations with
\begin{equation}
   \omega = k \sqrt{k^{2}+2\varrho} + U + \rho.
   \label{OmCondensateu}
\end{equation}
Formula (\ref{OmCondensateu})  is an important and nontrivial result 
which can be obtained neither from existing general facts about the 
WBK formalism nor from the linear theory of homogeneous systems.

%%%%%%%%%%%%%%%%%%%%%%%%%%%%%%%%%%%%%%%
\section{Weakly nonlinear GP equation }
%%%%%%%%%%%%%%%%%%%%%%%%%%%%%%%%%%%%%%%
The derivation for the description of the nonuniform turbulence found
in a BEC system consists of a amalgamation of a WKB method, for the
description of the linear dynamics, and a standard weak turbulence
theory (see e.g. \cite{DNPZ92}), with the noted modification that
Gabor transforms are used instead of Fourier ones.  We will now
demonstrate the general ideas of such a derivation for the simple case
of system where no condensate is present.

Consider the Gabor transformation of (\ref{eq1}):
\begin{equation}\label{eq1Gabor}
    i \PA{t} \hat \psi + \triangle \hat \psi 
                       - \widehat{|\psi|^{2}\psi} 
                       - U\hat\psi
                       + i {\bf  } (\nabla_x U)\nabla_k\hat{\psi}
                       = 0.
\end{equation}
To calculate the $\widehat{|\psi|^{2}\psi}$ term let us first separate
the Gabor transform into its correspondingly fast and slow spatial
parts,
\begin{equation} 
   \hat\psi(\vec{x},\vec{k},t) = \underbrace{a(\vec{x},\vec{k},t)}_{\mbox{slow}} 
                                 \underbrace{e^{i\kx}}_{\mbox{fast}}.
   \label{FastSlow}
\end{equation}
Now by using the inverse Gabor transform
\begin{equation}\label{GaborReverse}
   g(x,t)= \int \hat{g}(\vec{x},\vec{k},t) \, d {\vec k},
\end{equation}
we find 
\begin{align}
   \widehat{|\psi|^{2}\psi} = e^{i \kx } \int f&(\x-\x_0) \, e^{i \x_0
        \cdot (\k_3+ \k_2- \k_1- \k)} \nonumber \\ &\times a^*(\k_1, \x_0)
        a(\k_2, \x_0)a(\k_3, \x_0) \, d {\vec x_0} d {\vec k_1} d
        {\vec k_2} d {\vec k_3}. \nonumber \\ 
   \label{stuff1}
\end{align} 
Note that the slow amplitudes $a$ do not change much over the
characteristic width of the function $f$ and hence their argument
${\vec{x_0}}$ can be replaced by ${\vec{x}}$.  Therefore, we can
approximate (\ref{stuff1}) by
\begin{align}
   \widehat{|\psi|^{2}\psi} \simeq 
        \frac{e^{i \kx}}{(2 \pi)^{3 d/2}} \int 
        F&(\k_3+\k_2-\k_1-\k) \nonumber \\ 
        &\times a^*(\k_1,\x) a(\k_2,\x)a(\k_3,\x) 
        \, d {\vec k_1} d {\vec k_2} d {\vec k_3}. \nonumber \\ 
   \label{morestuff1}
\end{align}
Here $F(\k)$ is the Fourier transform of $f(\x)$. Note that for the
spatially homogeneous systems, $\epsilon^*\to 0$, $F(\k)$ is just a
delta function,
  $$\lim\limits_{ \epsilon^*\to 0}F(\k)\to \delta(\k).$$
After dropping terms proportional to $ \triangle a$, equation 
(\ref{eq1Gabor}) then becomes
\begin{align}
   \PA{t} a(\k,\x)  =  - 2 \k \cdot \nabla &a(\k,\x) \nonumber \\
                       - i(k^2 +\k \cdot(\nabla_x ))&a(\k,\x)
                       - (\nabla_x U) (\nabla_k a(\k,\x)) \nonumber \\
                       - \int F(\k_3 + \k_2 &- \k_1 - \k) 
                         \,a^*(\k_1,\x) a(\k_2,\x) a(\k_3,\x) 
                         \,d {\vec k_1} d {\vec k_2} d {\vec k_3}.
  \nonumber \\ 
   \label{morestuff2}
\end{align}
This is the master equation formulating the nonlinear dynamics in
terms of the Gabor amplitudes. This can serve as a starting point for
the statistical averaging which in turn leads to the weak turbulence
formalism. Note that this equation can be written in Hamiltonian form,
\begin{equation}
   i \frac{\partial}{\partial t} a_{{\bf k},{\bf x}}=
     \frac{\delta H} {\delta a_{{\bf x},{\bf k}}^*},
   \label{canonicalGfiltered7}
\end{equation}
with a Hamiltonian function
\begin{align}
  H   = \int 
        ( \omega_{k,x} - {\bf x} \cdot&\nabla_x \omega_{k,x} ) |a_{k,x}|^2 \nonumber \\  
      + \frac{i}{2} (\nabla_x &\omega_{k,x})
        ( a_{k,x}^{*}\nabla_{k}a_{k,x} - a_{k,x}\nabla_{k}a^{*}_{k,x} ) \nonumber\\ 
      + \frac{i}{2}&(\nabla_k \omega_{k,x}) 
        ( a_{k,x}\nabla_{x}a_{k,x}^{*} - a_{k,x}^{*}\nabla_{x}a_{k,x} ) 
        \, d {\bf k} d {\bf x}  \nonumber\\
      + \int 
         F({\bf k_3} + {\bf k_2}& - {\bf k_1} - {\bf k}) \nonumber \\
         a^{*}({\bf k_1}&,{\bf x}) a({\bf k_2},{\bf x}) 
         a({\bf k_3},{\bf x}) a({\bf k_4},{\bf x})  
         \, d {\vec k_1} d {\vec k_2} d {\vec k_3} d {\vec k_4},   
 \nonumber \\ 
\label{ham}
\end{align}
where $\omega_{k,x}= k^2 + U(x)$. In fact, such a Hamiltonian
description can be derived directly, in terms of the Gabor amplitudes,
from the Hamiltonian formulation of the original GP equation (see
Appendix B).

If a condensate is present in the system, one can also re-write the
equations in a Hamiltonian form with an identical quadratic part. That
is, with $a$ being replaced by the normal amplitude, and $\omega$ by
the frequency of waves, found in the presence of the condensate.  It
appears that the quadratic part of the Hamiltonian (\ref{ham}) is
generic in the WKB context.  Indeed, let us consider a typical
Hamiltonian for linear waves in weakly inhomogeneous media
\cite{papaLvov} expressed in terms of Fourier amplitudes $a_{\bf q_1}$
and $ a^*_{\bf q_1}$
\begin{eqnarray}
  {\cal H}=\int\Omega({\bf q_1}, {\bf q}) \, a_{\bf q_1} a^*_{\bf q_1} 
  \, d {\bf q} d {\bf q_1},
  \label{PapinHamiltonian}
\end{eqnarray}
with a hermitian kernel $\Omega({\bf q_1}, {\bf q})= \Omega({\bf q},
{\bf q_1})$ which is strongly peaked at ${\bf q}- {\bf q_1} =0 $.  As
we will show in a separate paper \cite{naz-lvov}, this Hamiltonian can
be represented in terms of the Gabor transforms as
\begin{align}
   H = \int ( \omega_{\k,\x} - {\bf x} \cdot &\nabla_x \omega_{\k,\x}
       ) |a_{k,x}|^2 \nonumber \\ + \frac{i}{2} ( \nabla_x
       &\omega_{\k,\x} ) ( a_{k,x}^*\nabla_{k}a_{k,x} -
       a_{k,x}\nabla_{k}a^*_{k,x} ) \nonumber \\ + \frac{i}{2}& (
       \nabla_k \omega_{\k,\x} ) ( a_{k,x}\nabla_{x}a_{k,x}^* -
       a_{k,x}^*\nabla_{x}a_{k,x} ) \, d {\bf k} d {\bf x},
\end{align}
where $a_{k x}$ are the Gabor coefficients, and $\omega_{\k \x}$ is
the position dependent frequency, related to $\Omega({\bf q}, {\bf
q_1})$ via
\BEA \omega_{\k,\x}=\int e^{- 2 i \q\cdot\x } \, \Omega(\k,\k+2\q) \,
   d \q.  \EEA
Actually, such an expression is a canonical form, even for a much
broader class of Hamiltonians that correspond to a significant class
of linear equations with coordinate dependent coefficients
\cite{naz-lvov}. That is,
\BEA {\cal H}=\int [A({\bf q_1}, {\bf q}) \, a_{\bf q_1} a^*_{\bf q}
  \, B({\bf q_1}, {\bf q}) \, a_{\bf q_1} a_{\bf - q} + c.c.]  \, d
  {\bf q} d {\bf q_1},
  \label{genHamiltonian}
\EEA
where functions $A$ and $B$ peaked at ${\bf q}- {\bf q_1} =0 $.

%%%%%%%%%%%%%%%%%%%%%%%%%%%%%%%%%%%%%%%%%%%%%%%%%%%%%%%
\section{Weak turbulence for inhomogeneous systems}
%%%%%%%%%%%%%%%%%%%%%%%%%%%%%%%%%%%%%%%%%%%%%%%%%%%%%%%
Now, by analogy with homogeneous weak turbulence, we define the
waveaction spectrum as
  $$n_{\k,\x} = \langle |a(\k,\x) |^2 \rangle /F(0),$$
where averaging is performed over the random initial phases.  Note
that this definition is slightly different to the usual definition of
the turbulence spectrum in homogeneous turbulence, i.e. the definition
constructed from Fourier transforms, $n_{\k} \, \delta(\k - \k') =
\langle a(\k) a(\k') \rangle$.  Indeed, a Gabor transform can be
viewed as a finite-box Fourier transform, where $\k = \k'$ in the
definition of the spectrum and one replaces $\delta(\k - \k')$ with
the box volume $F(0)$.

Multiplying (\ref{morestuff2}) by $a^*(\k,\x)$ and combining the
resulting equation with its complex conjugate, we get a generalization
of (\ref{phi}):
\begin{align}\label{generalizationofphi}
   D_t n_{\k,\x} = -2 \Im \int F&(\k+\k_1-\k_2-\k_3) \nonumber \\ 
         &\times\langle a^*(\k,\x) a^*(\k_1,\x) a(\k_2,\x) a(\k_3,\x)\rangle
         \, d {\bf k_1} d {\bf k_2} d {\bf k_3}, \nonumber \\ 
\end{align}
with $\Dt \equiv \PA{t} + \dot{\vec{x}}\cdot\nabla
+\dot{\vec{k}}\cdot\PA{k}$.  Note, that in the case of homogeneous
turbulence, using the random phase assumption, in the above equation,
would lead to the RHS becoming zero. This means that the nontrivial
kinetic equation appears only in higher orders of the nonlinearity.
For the inhomogeneous case, the nontrivial effect of the nonlinearity
appears even at this (second) order. This can be seen via a frequency
correction which, in turn, modifies the wave trajectories.  This
effect was considered by Zakharov et al \cite{zmrub} and it is
especially important in systems where such frequency corrections
result in modulational instabilities followed by collapsing events. In
our case the nonlinearity is ``defocusing'' and, therefore, such an
effect is less important. Indeed, in what follows we will neglect this
effect as, at sufficiently small ratios of the inhomogeneity and
turbulence intensity parameters, $\epsilon \ll \phi^2$, wave collision
events are a far more dominant process.

Let us introduce notations
   $$I^{k k_1}_{k_2 k_3} \equiv \langle 
     a^*(\k,\x) a^*(\k_1,\x) a(\k_2,\x) a(\k_3,\x)\rangle,$$
and 
   $$I^{k k_1 k_2}_{k_3 k_4 k_5}  \equiv \langle
     a^*(\k,\x) a^*(\k_1,\x) a^*(\k_2,\x) a(\k_3,\x) a(\k_4,\x)
     a(\k_5,\x)\rangle.$$ 
Then,  we have the following equation for the 4th-order moment,
\begin{align}
   D_t I^{k_1' k_2'}_{k_3' k_4'} = 
       i(\tilde\omega_{k'_1} + \tilde\omega_{k_2'} - \tilde\omega_{k_3'}
      -  \tilde\omega_{k_4'}) &I^{k'_1 k_2'}_{k_3' k_4'} \nonumber \\
      + \int \Big(I^{k_2 k_3 k_2'}_{k_1 k_3' k_4'} F(\k_1'+\k_1-\k_2&-\k_3) \nonumber \\
           + I^{k_1' k_2 k_3 }_{k_1 k_3' k_4'} F(\k_2'+\k_1&-\k_2-\k_3) \nonumber \\
           - I^{k_1'  k_2' k_1}_{k_4' k_2 k_3} F(\k_3'&+\k_1-\k_2-\k_3) \nonumber \\
           - I^{k_1' k_2' k_1}_{k_3' k_2 k_3} &F(\k_4'+\k_1-\k_2-\k_3) \, 
\Big)d k_1 d k_2 d k_3,\nonumber \\
\label{ef4m}
\end{align}
where we denote $\tilde\omega_{k}=k^2 + (\k \cdot \nabla_x U)$.  Note
that the first two terms on the RHS of this equation can be obtained
one from another by exchanging $\k_1'$ and $\k_2'$, whereas the last
two terms -- by exchanging $\k_3'$ and $\k_4'$.  To solve this
equation, one can use the {\it random phase assumption} which is
standard for the derivation of a weak {\it homogeneous} turbulence
theory and which allows one to express the 6th-order moment in terms
of the 2nd-order correlators. For homogeneous turbulence, the validity
of this assumption was examined by Newell et al 
\cite{NazarenkoNewell,Newell_ann}
 who showed that initially Gaussian turbulence
(characterized by random independent phases) remains Gaussian for the
energy cascade range whereas in the particle cascade range deviations
from Gaussianity grow toward low $k$ values. However, these deviations
remain small over a large range of $k$ for small initial amplitudes
and the random phase assumption can be used for these scales. Note
that the deviations from Gaussianity at low $k$ correspond to the
physical process of building a coherent condensate state. The results
of \cite{NazarenkoNewell,Newell_ann} obtained for homogeneous GP
turbulence will hold for trapped turbulence too because inhomogeneity
has a neutral effect on the phase correlations. Indeed, according to
the linear WKB equations the phases propagate unchanged along the
rays.  Thus we write
\begin{eqnarray}
   I^{123}_{456} \approx n_1 n_2 n_3 \Big(   
          F^{3}_{4}(F^{2}_{5}F^{1}_{6}
                  + F^{1}_{5}F^{2}_{6})
        + &F^{3}_{5}&(F^{2}_{6}F^{1}_{4}
                  + F^{2}_{4}F^{1}_{6}) \nonumber \\
        &+& F^{3}_{6} (F^{1}_{5}F^{2}_{4}
                  + F^{1}_{4}F^{2}_{5}) \Big),\nonumber \\
   \label{dN6def} 
\end{eqnarray}
here we have used the shorthand notations, $F^{1}_{2}\equiv F(0) \,
\delta(\k_1-\k_2)$ and $I^{123}_{456}=I^{k_1 k_2 k_3}_{k_4 k_5 k_6}$.
Using this expression in (\ref{ef4m}) we have
\begin{eqnarray}
   \frac{D}{D t} I^{k_1 k_2}_{k_3 k_4} = 
         &i&(\omega_{k_1}+\omega_{k_2}
         - \omega_{k_3}
         - \omega_{k_4}) I^{k_1 k_2}_{k_3 k_4} \nonumber \\
         &+& 2 \left(n_{k_3} n_{k_4}(n_{k_1} + n_{k_2}) 
                - n_{k_1} n_{k_2}(n_{k_3} + n_{k_4})\right). \nonumber 
\end{eqnarray}
Notice that the $\tilde \omega_k$ terms get replaced by $\omega_k$,
since the $(\k \cdot \nabla_x U)$ terms drop out on the resonant
manifold.  Let us integrate this equation over the period $T$ which is
less than both the slow WKB time $1/\epsilon$ and the nonlinear time
$1/\sigma^4$.  Then, one can ignore the time dependence in $n_k$ on
the RHS of the above equation and we can take \mbox{$\dot{k} = -
\nabla U = const$} on the LHS.

The resulting equation can be easily integrated along the
characteristics (rays) which in the limit $\omega T \to \infty$ gives
\begin{align}
   I^{k_1 k_2}_{k_3 k_4} = 
        - 2 [ n_{k_3} n_{k_4}(n_{k_1} n_{k_2}) 
                  - n_{k_1} n_{k_2}(n_{k_3}&+n_{k_4}) ] \nonumber \\ 
            \delta(\omega_{k_1}+\omega_{k_2} &- \omega_{k_3} - \omega_{k_4}).
   \label{FourOrderCorrelator}
\end{align}
Note that to derive a similar expression in the theory of {\em
homogeneous} weak turbulence one usually introduces an artificial
``dissipation'' to circumvent the pole and to get the correct sign in
front of the delta function (see e.g. \cite{ZLF}). The roots of this
problem can be found even at the level of the linear dynamics, where
the use of Laplace (rather than Fourier) transforms provides a
mathematical justification for the introduction of such a
dissipation. However, in our case there is no need for us to introduce
such a dissipation because inhomogeneity removes the degeneracy in the
system.
Substituting (\ref{FourOrderCorrelator}) into
(\ref{generalizationofphi}) we get the main equation describing weak
turbulence, the four-wave kinetic equation
\begin{eqnarray}
   \Dt n_{k} = 
       \frac{1}{\pi} \int &n_k& n_1 n_2 n_3 \left(
             \frac{1}{n_{k}} 
           + \frac{1}{n_{1}} 
           - \frac{1}{n_{2}}
           - \frac{1}{n_{3}} \right)
       \delta \left(\vec{k} +\vec{k}_1 - \vec{k}_2 - \vec{k}_3\right)   
 \nonumber \\  
    && \delta \left(\omega_k({\bf x}) + \omega_1({\bf x}) - \omega_2({\bf x}) - \omega_3({\bf x})\right) 
       \, d \vec{k}_1 d \vec{k}_2 d \vec{k}_3,            
    \nonumber \\  
\label{FourWaveKineticEquationForGP}
\end{eqnarray}
where,
\begin{align}
   &\Dt \equiv \PA{t} + \dot{\vec{x}}\cdot\nabla
                     + \dot{\vec{k}}\cdot\PA{k},    
   &\dot{\vec{x}} = \PA{k}\omega, \,\,\,\,\,\,\,
    \dot{\vec{k}} = -\nabla\omega.  \nonumber
\end{align}
We can see that the main difference between the kinetic equation for
inhomogeneous media and homogeneous turbulence
\cite{SV,ZMR85,DNPZ92,Newell68} is that the partial time derivative on
the LHS is replaced by the full time derivative along the rays.
Further, the frequency $\omega$ and spectrum $n$ are now functions not
only of the wavenumber but also of the coordinate.

The same is true for the case when the ground state condensate is
important for the wave dynamics \cite{DNPZ92}. The main interaction
mechanism now become three wave interactions, with the kinetic
equation
\begin{eqnarray}
   D_t n &=& \pi \int |V_{k k_1 k_2}|^2 \, f_{k12} \,
           \delta_{{{\bf k} - \bf{k_1}-\bf{k_2}}} \, 
           \delta_{\omega_{{\bf k}} -\omega_{{\bf{k_1}}}-\omega_{{\bf{k_2}}}} 
           d {\bf k}_{1} d {\bf k}_2 \, 
   \nonumber \\ 
         &-& 2\pi\int \, 
           |V_{k_1 k k_2}|^2\, f_{1k2}\, 
           \delta_{{{\bf k_1} - \bf{k}-\bf{k_2}}} \,
           \delta_{{\omega_{{\bf k_1}} -\omega_{{\bf{k}}}-\omega_{{\bf{k_2}}}}}
        \, d {\bf k}_{1} d {\bf k}_2 \, , 
   \label{KE3wave}
\end{eqnarray}
where $ f_{k12} = n_{{\bf k_1}}n_{{\bf k_2}} - n_{{\bf k}}(n_{{\bf
k_1}}+n_{{\bf k_2}}) \,$. Here, $n_k$, $D_t$ and $\omega$ are given by
expressions (\ref{waccondenu}), (\ref{fdar}) and (\ref{OmCondensateu})
respectively and the expression for the interaction coefficient $V_{k
k_1 k_2}$ can be found in \cite{DNPZ92}. Three-wave interactions
always dominate over the four-wave process when $\rho \sim k^2 $
(because $k \sim 1$ and $n \ll 1)$.  In the case $\rho \ll k^2 $, the
relative importance of the three-wave and the four-wave processes can
be established by comparing the characteristic times associated with
these processes.  The characteristic time of the three wave
interactions for $\rho \ll k^2 $ is
\begin{equation*}
   \tau_{3w} = k^{2-d}/\varrho n.
\end{equation*}
Thus, the 3-wave process will
dominate the 4-wave one if the condensate is stronger than the waves, i.e. if 
$\varrho > n k^d \sim \phi^2$.

%%%%%%%%%%%%%%%%%
\section{Summary}
%%%%%%%%%%%%%%%%%
In this paper, we developed a theory of weak inhomogeneous wave
turbulence for BEC systems. We started with the GP equation and
derived a statistical theory for the BEC kinetics which, in
particular, describes states which are very far from the thermodynamic
equilibrium.  Such nonequilibrium states take the form of wave
turbulence which is essentially inhomogeneous due to the fact that the
BEC is trapped by an external field.  There are two main new results
in this paper. First of all, we have described the effect of the
inhomogeneous ground state on the linear wave dynamics and, in
particular, we have shown that such an effect cannot be modeled by
renormalizing the trapping potential as it was previously suggested in
literature.  This was done by deriving a consistent WKB theory based
on the scale separation between the ground state and the waves. Our
results show that the condensate ``mildly'' pushes the wave turbulence
away from the center but it can never reflect it (as an external
potential would). Note that we established this result only for the
limit of large occupation numbers described by the GP equation and
this, in principle, does not rule out a possibility that the the
renormalized potential approach can still be valid in the opposite
limit of small occupation numbers.  Secondly, we showed that the
kinetic equation for trapped waves generalizes, and one can combine
the linear WKB theory and the theory of homogeneous weak turbulence in
a straightforward manner. Namely, the partial time derivative on the
LHS of the kinetic equation is replaced by the full time derivative
along the wave rays, while the frequency and the spectrum on the RHS
now become functions of coordinate. A suitable definition for the
coordinate dependent spectrum is given by using the Gabor transforms
instead of Fourier transforms. It is important to notice that the
coordinate dependence of the wave frequency has a profound effect on
the nonlinear dynamics. The resonant wave interactions can now take
place only over a limited range of wave trajectories which makes such
interactions similar to the collision of discrete particles.

Similarly to the case of homogeneous turbulence considered in
\cite{DNPZ92}, the presence of a condensate changes the resonant wave

interactions from four-wave to three-wave if the condensate intensity
exceeds that of the waves. A distinct feature of the inhomogeneous
turbulence trapped by a potential is that if the three-wave regime is
dominant in the center of the potential well, it is likely to be
suddenly replaced by a four-wave dynamics when one moves out of the
center beyond the condensate reflection points where the condensate
intensity is decaying exponentially fast.  Thus the same wavepacket
can alternate between three-wave and four-wave interactions, with
other wavepackets, as it travels back and forth between its reflection
points in the potential well. (The wavepacket reflection points being
further away from the center than the condensate's own reflection
points).
%%%%%%%%%%%%%%%%%%%%%%%%%%%%%%%%%%%%%%%%%%%%%%%%%%%%%%%%%%%%%%%%%%%%%%%%%%%%
\section*{Appendix A: derivation WKB equations in presence of a condensate }
%%%%%%%%%%%%%%%%%%%%%%%%%%%%%%%%%%%%%%%%%%%%%%%%%%%%%%%%%%%%%%%%%%%%%%%%%%%%
Let us split $\phi$ into its real and imaginary parts $a = \Re \phi$
and $b = \Im \phi $.  Then the equation (\ref{eq3}) splits into two
coupled equations
\begin{align}
   \PA{t}a &+ \triangle b + 2\vec{v}\cdot\nabla a
            +\frac{\nabla\varrho}{\varrho}\cdot\nabla b 
            +\rho(2 a b + b(a^2+b^2))
            = 0, \label{eq5} 
\\ \PA{t}b &- \triangle a +2a\varrho+2\vec{v}\cdot\nabla b
            -\frac{\nabla\varrho}{\varrho}\cdot\nabla a 
            + {\varrho}(3 a^2+b^2 +a(a^2+b^2))
            = 0, \label{eq6}
\end{align}
where we have used the fact that
\mbox{$ \frac{\nabla\psi_{0}}{\psi_{0}} =
\frac{\nabla\varrho}{2\varrho}+\frac{i}{2} \vec{v} $}, which follows from
(\ref{condensaterepresentation}).
                                                                          
Gabor transforming our two coupled equations (\ref{eq5}) and
(\ref{eq6}) and using Taylor series to represent large-scale
quantities,
\begin{equation*}
   \varrho(\vec{x}_{0}) = \varrho(\vec{x}) +
    (\vec{x}_{0}-\vec{x})\cdot\nabla\varrho(\vec{x}) +
   O(\varepsilon^{2}),
\end{equation*}
we find
\begin{align}
     \PA{t}\hat{a} + \triangle \hat{b} 
                   + \frac{\nabla \varrho}{\varrho}\cdot \nabla \hat{b}
                   + \vec{v}\cdot\nabla\hat{a} 
                  &+{\cal G} \left[\rho(2 a b + b(a^2+b^2))\right]
                   = 0, \label{eq7}\\
     \PA{t}\hat{b} - \triangle \hat{a} 
                   - {\bf  }\frac{\nabla \varrho}{\varrho}\cdot \nabla \hat{a}
                   + 2{\bf  }\vec{v}\cdot\nabla\hat{b} 
                  &+ 2\varrho\hat{a} 
                   + 2i{\bf  }\nabla\varrho\cdot\PA{k}\hat{a} \nonumber \\
                  &+ {\cal G}\left[( \varrho(3 a^2+b^2 +a(a^2+b^2)))\right] 
                   = 0\label{eq8}.
\end{align}
Where ${\cal G}[f(x)]$ is the Gabor transform of $f(x)$. We have kept
only $O(\varepsilon)$ terms and neglected the $O(\varepsilon^{2})$ and
higher order terms. For generality, we have kept the nonlinear term.

%----------------------------------------%
\subsubsection{The $\epsilon^0$ order - }
%----------------------------------------%
As in all WKB based theories we first derive a linear dispersion
relationship from the lowest order terms.  At zeroth order in ${\bf
\epsilon}$, the spatial derivative of a Gabor transform is
$\nabla\hat{a} = i \vec{k}\hat{a}$ which is similar to the
corresponding rule in Fourier calculus. Then, at the lowest order,
equations (\ref{eq7}) and (\ref{eq8}) become
\begin{align}
   \PA{t}\hat{a} &- k^{2} \hat{b} =0, \label{eq9}\\
   \PA{t}\hat{b} &+ k^{2} \hat{a} + 2\varrho\hat{a}
   =0.\label{eq10}
\end{align}
These two linear coupled equations make up an eigenvalue
problem. Diagonalizing these equations we obtain
\begin{align*}
    &\PA{t}\lambda = +i\omega\lambda,  &\PA{t}\mu = -i \omega\mu.
\end{align*}
 Correspondingly, we find the eigenvectors
\begin{align}\label{eq11}
    &\lambda = \frac{1}{2}\left(\hat{a}
             - \frac{i k^{2}}{\omega}\hat{b}\right),
    &\mu = \frac{1}{2}\left(\hat{a}
         + \frac{i k^{2}}{\omega}\hat{b}\right),
\end{align}
 or, re-arranging for $\hat{a}$ and $\hat{b}$
\begin{align}\label{eq12}
    &\hat{a} = \lambda + \mu,
    &\hat{b} = \frac{i\omega}{k^{2}}(\lambda-\mu).
\end{align}
The eigenvalues are given by the dispersion relationship,
\begin{equation}
    \omega^{2} = k^{2}(k^{2}+2\varrho),
    \label{BogolubovDispersion}
\end{equation}
which is identical to the famous Bogoliubov
form~\cite{Bogolubov1947} which was also obtained for  waves on a 
homogeneous condensate in the weak turbulence context in \cite{DNPZ92}.

Therefore, at the zeroth order, we see that $\lambda$ rotates with
frequency $-\omega$ and $\mu$ rotates at $+\omega$. Note that the
$\lambda$ and $\mu$ are related via
\begin{equation}
   \lambda^{*}(\vec{k}) = \mu (-\vec{k}) 
   \label{eq14}.
\end{equation}
%

%---------------------------------------% 
\subsubsection{The $\epsilon^1$ order -}
%---------------------------------------%
%
Let us split the wave amplitudes into fastly and
slowly  varying parts,
\begin{align}\label{fastslow}
   &\lambda(\vec{x},\vec{k},t) = 
    \Lambda(\vec{x},\vec{k},t)\E^{i \vec{k}\cdot\vec{x}+i \omega t},
   &\mu (\vec{x},\vec{k},t) = 
     M(\vec{x},\vec{k},t)\E^{i \vec{k}\cdot\vec{x}-i \omega t},
\end{align}
or, in shorthand notation,
\begin{align}\label{fastslow2}
   &\lambda = \Lambda\E^{+},
   &\mu = M\E^{-},
\end{align}
where
\begin{align*}
   &\E^{+} \equiv \E^{i \vec{k}\cdot\vec{x}+i \omega t},
   &\E^{-} &\equiv \E^{i \vec{k}\cdot\vec{x}-i \omega t}.
\end{align*}
The $\E^{+}$ and $\E^{-}$ represent the fastly oscillating parts of
the Gabor transforms.   From (\ref{eq14}) it follows that
\begin{equation}
   \Lambda^*(\vec{x},-\vec{k},t) = M(\vec{x},\vec{k},t).
   \label{lm-rel}
\end{equation}
Obviously,
\begin{align}\label{result1}
   &\PA{t}\lambda = i \omega\lambda + \E^{+}\PA{t}\Lambda,
   &\PA{t}\mu = -i \omega\mu + \E^{-}\PA{t}M,
\end{align}
\begin{align}\label{result2}
   &\nabla\lambda = i \vec{k}\lambda + \E^{+}\nabla\Lambda 
                                     + i  t\lambda\nabla\omega,
   &\nabla\mu = i \vec{k}\mu + \E^{-}\nabla M -i  t\mu\nabla\omega,
\end{align}
\begin{align}\label{result3}
   \triangle \lambda &= -k^{2}\lambda 
                        + 2i  \E^{+}\vec{k}\cdot\nabla\Lambda 
                        + \E^{+}\triangle\Lambda 
                        - 2t\lambda\vec{k}\cdot\nabla\omega,\\ 
   \triangle \mu &= -k^{2}\mu 
                        + 2i  \E^{+}\vec{k}\cdot\nabla M 
                         + \E^{+}\triangle M 
                         + 2t\mu\vec{k}\cdot\nabla\omega, \nonumber
\end{align}
\begin{align}\label{result4}
   &\PA{k}\lambda = \E^{+}\PA{k}\Lambda 
              + i \vec{x}\E^{+}\Lambda + i  t\E^{+}\Lambda\PA{k}\omega,\\
   &\PA{k}\mu = \E^{-}\PA{k}M 
              + i \vec{x}\E^{-}M - i  t\E^{-}M\PA{k}\omega. \nonumber
\end{align}

Our aim now is to derive equations for $\PA{t}\lambda$ and
$\PA{t}\mu$. However, due to the relationship (\ref{eq14}) it is
sufficient to derive an equation for only one of the two, for example
$\lambda$. From (\ref{eq11}) we find
\begin{equation*}
    \PA{t}\lambda = \PA{t}\left(\frac{\hat{a}}{2} 
                  - \frac{i k^{2}}{2\omega}\hat{b}\right).
 \end{equation*}
After substituting our equations for $\PA{t}\hat{a}$ and
$\PA{t}\hat{b}$, (\ref{eq7}) and (\ref{eq8}), and making use of the
relationships (\ref{eq12}) the equation for $\lambda$ acquires the
following form:
\begin{align*}
   \PA{t}\lambda &= \lambda \left[-\frac{i \nabla\varrho\cdot\nabla\omega}{2k^{2}\varrho} 
                  + \frac{i k^{2}\varrho}{\omega}\right]
                  + \nabla\lambda \cdot \left[-\frac{i \nabla\omega}{k^{2}}
                  - \frac{i \omega\nabla\varrho}{2k^{2}\varrho}
                  - 2\vec{v}
                   - \frac{i k^{2}\nabla\varrho}{2\omega\varrho}\right] \\
                  &+ \triangle\lambda \left[-\frac{i \omega}{2k^{2}}
                   -\frac{i k^{2}}{2\omega}\right]
                   -\frac{k^{2}}{\omega} \nabla\varrho\cdot\PA{k}\lambda\\
                  &+ \mu \left[\frac{i \nabla\varrho\cdot\nabla\omega}{2k^{2}\varrho} 
                   + \frac{i k^{2}\varrho}{\omega}\right]
                   + \nabla\mu \cdot \left[+\frac{i \nabla\omega}{k^{2}}
                   + \frac{i \omega\nabla\varrho}{2k^{2}\varrho}
                   - \frac{i k^{2}\nabla\varrho}{2\omega\varrho}\right] \\
                 &+ \triangle\mu \left[+\frac{i \omega}{2k^{2}}
                  -\frac{i k^{2}}{2\omega}\right]
                  -\frac{k^{2}}{\omega}\nabla\varrho\cdot\PA{k}\mu  - {\cal{NL}}.
\end{align*}  
Here the nonlinear term ${\cal{NL}}$ is given by
  $$ {\cal{NL}}={\cal G} \left[\rho(2 a b + b(a^2+b^2))\right] -\frac{i
      k^2}{2\omega_k} {\cal G}\left[( \varrho(3 a^2+b^2+a(a^2+b^2)))\right].$$
Note that we have neglected $\dot \omega$ in the above expressions
because, according to the dispersion relationship
(\ref{BogolubovDispersion}), it is of the order of $\dot \rho$ which
is $O(\epsilon^2)$ by virtue of (\ref{continuity}). We will also drop
the nonlinear term in the subsequent calculation.

Our next step is to eliminate the fast oscillations associated with
the Gabor transforms and derive an equation for $|\Lambda|^{2}$. This
in turn will lead to a natural waveaction quantity which can be used
to describe the behavior of our wavepackets in phase space. Using
(\ref{result1}-\ref{result4}) we obtain
\begin{align}
    \PA{t}\Lambda &= \Lambda \left[-\frac{i \nabla\varrho\cdot\nabla\omega}{2k^{2}\varrho}
                   + \frac{i k^{2}\varrho}{\omega} -i \omega \right]\nonumber\\
                  &+ \left[i \vec{k}\Lambda+\nabla\Lambda+i  t\Lambda\nabla\omega\right]
                     \cdot\left[-\frac{i \nabla\omega}{k^{2}}
                   - \frac{i \omega\nabla\varrho}{2k^{2}\varrho}
                   - 2\vec{v}
                   - \frac{i k^{2}\nabla\varrho}{2\omega\varrho}\right]\nonumber\\
                  &+ \left[-k^{2}\Lambda+2i \vec{k}\cdot\nabla\Lambda-2t\Lambda\vec{k}\cdot\nabla\omega\right]
                     \left[-\frac{i \omega}{2k^{2}}-\frac{i \vec{k}^{2}}{2\omega}\right]\nonumber\\
                  &-\frac{k^{2}}{\omega}\nabla\varrho\cdot\PA{k}\Lambda
                   -\frac{i k^{2}\Lambda}{\omega}\vec{x}\cdot\nabla\varrho
                   -\frac{i  tk^{2}\Lambda}{\omega}\nabla \varrho
\cdot\PA{k}\omega.\nonumber \\
\label{big1}
\end{align}
Please note that all the terms involving $M$ drop out. This stems from
the fact that, in deriving an equation for $\Lambda$, we have had to
divide through by $\E^{+}$. Therefore, any terms involving $M$ will
result in a factor
\begin{equation*}
  \E^{-}/\E^{+} = \E^{-2i \omega t}. 
\end{equation*}
Thus, after time averaging over a few wave periods, 
%denoted by the overline, 
all the $M$ terms drop out.

Expanding out equation (\ref{big1}) we find the $O(1)$ terms cancel
out and using the dispersion relationship (\ref{BogolubovDispersion}) we find

\begin{equation}\label{big2}
    \PA{t}\Lambda = \PA{k}\omega\cdot\nabla\Lambda +
                   \frac{\Lambda\omega}{\varrho
                   k^{2}}\vec{k}\cdot\nabla\varrho -
                   \nabla\omega\cdot\PA{k}\Lambda + i J ,
\end{equation}
where
\begin{equation*}
                J  = \frac{t\Lambda\omega}{k^{2}}\vec{k}\cdot\nabla\omega
                   + \frac{t\vec{k}^{2}\Lambda}{\omega}\vec{k}\cdot\nabla\omega
                   - 2\Lambda\vec{k}\cdot\vec{v}
                   - \frac{k^{2}\Lambda}{\omega}\vec{x}\cdot\nabla\varrho
                   - \frac{tk^{2}\Lambda}{\omega}\nabla\varrho\cdot\PA{k}\omega,
\end{equation*}
At this point let us drop the nonlinear term and concentrate on the
linear dynamics. Multiplying (\ref{big2}) by $\Lambda^{*}$ and
combining it with the complex conjugate equation the $J$ terms cancel,
leading to
\begin{equation}
    \PA{t}|\Lambda|^{2} - \PA{k}\omega\cdot\nabla|\Lambda|^{2} 
                        + \nabla\omega\cdot\PA{k}|\Lambda|^{2}    
                        = \frac{2|\Lambda|^{2}\omega}
                        {\varrho k^{2}}\vec{k}\cdot\nabla\varrho.
    \label{l-eq}
\end{equation}
A similar equation for $|M|^{2}$ can be easily obtained by replacing
${\bf k} \to {- \bf k}$ in (\ref{l-eq}) and using (\ref{lm-rel}),
\begin{equation}
    \PA{t}|M|^{2} + \PA{k}\omega\cdot\nabla|M|^{2} 
                  - \nabla\omega\cdot\PA{k}|M|^{2}    
                  = - \frac{2|M|^{2}\omega}
                  {\varrho k^{2}}\vec{k}\cdot\nabla\varrho.
    \label{m-eq}
\end{equation}
The LHS of this equation is the full time derivative of $|M|^{2}$
along trajectories.  If $|M|^{2}$ were to be a correct phase-space
waveaction, the right hand side of this equation would be zero,
however, this is not the case. We find the correct waveaction
$n(\vec{x},\vec{k},t) $ by setting
\begin{equation*}
   |M|^{2} = \alpha(\vec{x},\vec{k})n(\vec{x},\vec{k},t),
\end{equation*}
and finding such $\alpha(\vec{x},\vec{k})$ that the the full time
derivative of $n(\vec{x},\vec{k},t)$ is zero.  This leads to the
following condition on $\alpha$,
\begin{equation*}\label{ExtraJunk}
    \PA{k}\omega\cdot\nabla\alpha 
  - \nabla\omega\cdot\PA{k}\alpha 
  + \frac{2\alpha\omega}{\varrho k^{2}}\vec{k}\cdot\nabla\varrho = 0.
\end{equation*}
By choosing $\alpha = k^{X}\varrho^{Y}$ and substituting it to
(\ref{ExtraJunk}) we find $x=2$, $y=-1$.  Therefore the correct form
of the waveaction is $n = \frac{\varrho}{k^{2}}|M|^{2}.$ Summarizing,
we have got the following transport equation for the waveaction $n$ in
the linear approximation,
\begin{equation}
   \Dt n(\vec{x},\vec{k},t) = 0,
   \label{WCondensateWKB1}
\end{equation}
where
\begin{equation}
   \Dt \equiv \PA{t} + \dot{\vec{x}}\cdot\nabla 
                     + \dot{\vec{k}}\cdot\PA{k},
\end{equation}
is the full time derivative 
along trajectories and 
\begin{equation}
    \dot{\vec{x}} = \PA{k} \omega, \hspace{1cm}
    \dot{\vec{k}} = -\nabla \omega,
\end{equation}
are the ray equations with
\begin{equation}
    \omega = k \sqrt{k^{2}+2\varrho}.
    \label{OmCondensate1}
\end{equation}
Obviously, the dynamics in this case cannot be reduced to the
\EM{Ehrenfest} theorem with any shape of potential $U$. Therefore,
approaches that model the condensate effect by introducing a
renormalized potential are misleading.

Finally, it is useful to express the waveaction $n$ in terms of
the original variables,
\begin{equation} 
   n({\bf k},x,t) = \frac{1}{2}
                    \frac{\omega\rho}{k^{2}} \left| \widehat{ \Re \phi}
                  - \frac{ i k^{2} }{\omega} \widehat{ \Im \phi} \right|^{2}.
   \label{wacconden1}
\end{equation}
It is interesting that such a waveaction is in agreement with that found in
\cite{DNPZ92}. In fact in \cite{DNPZ92} the homogeneous case with non-zero nonlinearity
($\varepsilon = 0$, $\sigma \neq 0 $) was considered. This is the opposite limit 
to the one we have considered above (where $\varepsilon \neq 0$, $\sigma = 0$).

%%%%%%%%%%%%%%%%%%%%%%%%%%%%%%%%%%%%%%%%%%%%%%%%%%%%%%%%%%%%%%%%%%%%%%%%%%%%%%%%%%%%%%%%%
\section*{Appendix B: Hamiltonian formalism for spatially inhomogeneous weak turbulence.}
%%%%%%%%%%%%%%%%%%%%%%%%%%%%%%%%%%%%%%%%%%%%%%%%%%%%%%%%%%%%%%%%%%%%%%%%%%%%%%%%%%%%%%%%%

Let us start with the GP equation written in the Hamiltonian form:
\begin{equation}
   i\frac{\partial}{\partial t} \Psi_{\bf x} = 
    \frac{\delta {\cal H}}{\delta \Psi_{\bf x}^*}.\label{canonical}
\end{equation}
The Hamiltonian for the GP equation (\ref{eq1}) coincides with the
total energy of the system:
\begin{equation}
   {\cal H} = \int d{\bf r}\left( |\nabla \Psi_{\bf x}|^2
            + \frac{1}{2}|\Psi_{\bf x}|^4 
            + U({\bf x}) |\Psi_{\bf x}|^2 \right).
   \label{GPHam}
\end{equation}
Let us first consider the case without a condensate.  Applying the
Gabor transformation to (\ref{canonical}) we get
\begin{equation}
   i\frac{\partial}{\partial t} \hat\Psi_{\bf x} 
   = \widehat{\frac{\delta {\cal H}}{\delta \Psi_{\bf x}^*}}.
\end{equation}
But if we notice that 
$$ \frac{\delta {\cal H}(\Psi) }{\delta \Psi_{\bf x}} =
\frac{\delta {\cal H}(\hat\Psi) }{\delta \widehat{\Psi_{{\bf
x},{\bf k}}} },$$ we obtain
\begin{equation}
   i\frac{\partial}{\partial t} \hat\Psi_{\bf x} = 
    \widehat{\frac{\delta {\cal H}}{\delta\widehat{\Psi_{{\bf x},
    {\bf k}}^*}}}.
   \label{canonicalG}
\end{equation}
Thus, the time evolution of the Gabor transformed quantity is governed
by the Gabor transformed Hamiltonian equation. However, we would like
to obtain the equation of motion in Hamiltonian form without the Gabor
transformation. Let us re-write (\ref{canonicalG}) in terms of the
slow amplitudes $a$ defined in (\ref{FastSlow})
\begin{equation}
   i \frac{\partial}{\partial t} a_{{\bf k},{\bf x}} =
     \int f({\bf x} - {\bf x'}) \frac{\partial {\cal H}}{\partial a_{
           {\bf x'},{\bf k}}^*}d {\bf x'}.
   \label{canonicalGf}
\end{equation}
Now, let us express the Hamiltonian
(\ref{GPHam}) in terms of the slow variables
$a$,
\begin{eqnarray}
   {\cal H} &=& \int 
              e^{i(k_1 - k_2)x}\,
              \left( -a_{\bf k_1, x}(-k_2^2 + 2 i k_2 \nabla) 
              a_{\bf k_2, x}^* +U({\bf x}) 
              a_{\bf k_1, x} a_{\bf k_2, x}^* \right)\,
              d{\bf r} d{\bf k_1} d {\bf k_2}\nonumber\\
            && + \int 
              e^{i(k_1 + k_2-k_3 - k_4)x}\,
              a_{\bf k_1, x} a_{\bf k_2, x} a_{\bf k_3, x}^* a_{\bf k_4, x}^*
               \, d{\bf r} d{\bf k_1} d {\bf k_2} d {\bf k_3} d {\bf k_4}.
\nonumber\\   
\label{GPHam2}
\end{eqnarray}
Here we have integrated by parts $|\nabla \Psi|^2$ and, while
calculating the Laplacian of $\Psi$ in terms of slow variables, have
kept only the first order gradients in $a_{k,x}$. Substituting
(\ref{GPHam2}) into (\ref{canonicalGf}) allows us to re-write this
equation as
\begin{equation}
   i \frac{\partial}{\partial t} a_{{\bf k},{\bf x}} =
     \frac{\delta H}{\delta a_{
     {\bf x},{\bf k}}^*},\label{canonicalGfiltered2}
\end{equation}
where the filtered Hamiltonian $H$ can be
represented as
\begin{align}
   H = \int 
       f(&x-x')\,e^{i(k_1 - k_2)x'} \nonumber \\
       &\times\left( -a_{\bf k_1, x'} (k_2^2+ 2 i k_2 \nabla)a_{\bf k_2, x}^* 
               + U({\bf x'}) a_{\bf k_1, x'} a_{\bf k_2, x}^* \right)
       \,d{\bf r} d{\bf r'} d{\bf k_1} d {\bf k_2} \nonumber\\
     + \int 
       F(&{\bf k_1}+{\bf k_2 }-{\bf k_3 }-{\bf k_4 })\,
       a_{\bf k_1, x} a_{\bf k_2, x} a_{\bf k_3, x}^* a_{\bf k_4, x}^*
       \,d{\bf x} d{\bf k_1} d {\bf k_2} d {\bf k_3} d {\bf k_4},
\nonumber\\   
\label{GPHamfiltered2}
\end{align}
and $F({\bf k})$ is the Fourier transform of the $f(\x)$.

Expanding $U(x')$ as $U(x)+(x-x')\nabla U(x)$ and taking into account that 
$(x-x')$ can be interpreted as $-i\partial_{k_2} e^{i k_2 (x-x')}$,
we have
\begin{align}
   H  = \int\Big( 
          ( k^2 &+ U(x) ) |a_{k, x}|^2 
      - \frac{i}{2} (\nabla U(x))\hat\Psi_{k, x}\partial_{k}\hat\Psi_{k, x}^* 
      +  i k a_{k, x}\nabla a_{k, x}^* \nonumber \\
     &+  \frac{i}{2} (\nabla U(x))\hat\Psi_{k, x}^*\partial_{k}\hat\Psi_{k, x} 
      -  i k a_{k, x}^*\nabla a_{k, x}\Big)\, d {\bf k} d {\bf x} \nonumber \\ 
      + \int 
          F(&{\bf k_1}+{\bf k_2 }-{\bf k_3 }-{\bf k_4 })\,
          a_{\bf k_1, x} a_{\bf k_2, x} a_{\bf k_3, x}^* a_{\bf k_4, x}^*\,
     d{\bf x} d{\bf k_1} d {\bf k_2} d {\bf k_3} d {\bf k_4},\nonumber
\end{align}
Since $\omega_{k,x}= k^2 + U(x)$ we can represent the above 
formula as 
\begin{align}
  H = \int \Big(
      (\omega_{k, x} &- x \nabla \omega_{k, x}) |a_{k, x}|^2
    + \frac{i}{2}(\nabla_x \omega_{k x}) 
      \left(
        a_{k, x}^*\nabla_{k}a_{k, x} - a_{k, x}\nabla_{k}a^*_{k,x} 
      \right) 
      \nonumber\\ 
    &+ \frac{i}{2}(\nabla_k \omega_{k, x})
      \left( 
        a_{k, x}\nabla_{x}a_{k, x}^* - a_{k, x}\nabla_{k}a^*_{k,x} 
      \right)\Big) \,d {\bf k} d {\bf x}  \nonumber\\ 
    + \int 
       F({\bf k_1}&+{\bf k_2 }-{\bf k_3 }-{\bf k_4 })\,
       a_{\bf k_1 x} a_{\bf k_2 x} a_{\bf k_3 x}^* a_{\bf k_4 x}^*\,
       d{\bf x} d{\bf k_1} d {\bf k_2} d {\bf k_3} d {\bf k_4},
\nonumber\\ 
\label{favoritH}
\end{align}

Now, we will show that if a condensate is present then the quadratic
part of the Hamiltonian can also be written in the same canonical form
as in (\ref{favoritH}).  Let us start from the equation (\ref{big2})
for $\Lambda$
\begin{equation}\label{big2COPY}
    \PA{t}\Lambda  = \PA{k}\omega\cdot\nabla\Lambda 
                   + \frac{\Lambda\omega}{\varrho k^{2}}\vec{k}\cdot\nabla\varrho
                   - \nabla\omega\cdot\PA{k}\Lambda + i  J ,
\end{equation}
with 
\begin{equation*}
                J  = \frac{t\Lambda\omega}{k^{2}}\vec{k}\cdot\nabla\omega
                   + \frac{t\vec{k}^{2}\Lambda}{\omega}\vec{k}\cdot\nabla\omega
                   - 2\Lambda\vec{k}\cdot\vec{v}
                   - \frac{k^{2}\Lambda}{\omega}\vec{x}\cdot\nabla\varrho
                   - \frac{tk^{2}\Lambda}{\omega}\nabla\varrho\cdot\PA{k}\omega,
\end{equation*}
Expression (\ref{wacconden1}) for the waveaction in this case allows us to guess the
form of the normal variable,
$$a_{\k,\x} = \frac{\sqrt{\varrho\omega_{\k,\x}}}{k} \Lambda_{\k,\x}\,e^{i\omega_{\k,\x}}.$$
Note that this expression is consistent with the waveaction considered
above for the case with no condensate. This can be checked by taking
the limit $\rho \to 0$.  In terms of normal variable $a_{\k,\x} $
equations (\ref{big2}) and (\ref{big2COPY}) acquire the following
form:
\begin{align}
  \dot a_{\k,\x} = i \omega_{\k,\x} a_{\k,\x} &+ \partial_{\k}\omega_{\k,\x}\nabla a_{\k,\x} 
                 + \nabla \omega_{\k,\x} \partial_{\k}  a_{\k,\x} \nonumber \\
                 &- 2 i  a_{\k,\x}\k\cdot{\vec v} - i  a_{\k,\x}\x \cdot\nabla\omega_{\k,\x}.\nonumber\\ 
\label{EquationOfTransport}
\end{align}
This equation can be represented in the form of a Hamiltonian equation
of motion with a quadratic Hamiltonian as in (\ref{favoritH}) when
the frequency is replaced by its Doppler shifted value,
$$ \omega \to \omega + 2 \k\cdot{\vec v}. $$ Note that the Doppler
shift does not enter into the equation for the waveaction because it
leads to terms that are of second order in $\epsilon$ and therefore
should be neglected.

%%%%%%%%%%%%%%%%%%%%%%%%%%%%%%%%%%%%%%%%%%%%%%%%%%%%%%%%%
%%%%%%%%%%%%%%%%%% THE BIBLIOGRAPHY %%%%%%%%%%%%%%%%%%%%%
%%%%%%%%%%%%%%%%%%%%%%%%%%%%%%%%%%%%%%%%%%%%%%%%%%%%%%%%%


\begin{thebibliography}{10.}
\addcontentsline{toc}{section}{References}
\bibitem{Anderson} M.H. Anderson {\it et al.} Science {\bf 269}, 198
(1995).
\bibitem{Bradley} C.C.Bradley {\it et al.}, Phys.Rev.Lett, {\bf 75},
1687, (1995).
\bibitem{Davis} K.B.Davis {\it et al.}, Phys.Rev.Lett, {\bf 75} 3969,
(1995).
\bibitem{Bose} Bose S.N., Z.Phys., {\bf 26}, 178, (1924).
\bibitem{Einstein} Einstein A., Sitzber. Kgl. Preuss.Akad Wiss., 261 (1924); 3 (1925).
\bibitem{Pitaevsky} Dalfovo F., Giorgini S., Pitaevsky L., Stringari
S., Review of Modern Physics, {\bf 71}, 463, (1999).
\bibitem{Gross} E.P.Gross, {\it Nuovo Cimento}, {\bf 20}, 454, (1961).
\bibitem{Pitaevsky1961} L.P.Pitaevsky, {\it Sov. Phys. JETP}, {\bf
13}, 451 (1961).
\bibitem{Gardiner} C.W.Gardiner {\it et al.}, Phys.Rev.Lett, {\bf 81},
5266, (1998).
\bibitem{Gardiner2} C.W.Gardiner {\it et al.}, Phys.Rev.Lett, {\bf
79}, 1793, (1998).
\bibitem{SV} Kagan Yu.M, Svistunov B.V. and Shlyapnikov G.P.,
Sov. Phys. JETP, {\bf 75}, 387, (1992).
\bibitem{ZMR85} V.E. Zakharov, S.L. Musher, A.M. Rubenchik:
Phys. Rep. \textbf{129}, 285 (1985)
\bibitem{DNPZ92} S. Dyachenko, A.C. Newell, A. Pushkarev,
V.E. Zakharov: Physica D \textbf{57}, 96 (1992)
\bibitem{ZLF}V.E.  Zakharov, V.S.  L'vov 
and G.Falkovich, "Kolmogorov Spectra of Turbulence", Springer-Verlag, 1992.
\bibitem{Newell} B.  J . Benney and A.C.  Newell, Studies in Appl.
Math.  {\bf 48} (1) 29 (1969).
\bibitem{Ben} D.J. Benney and P.Saffman, Proc Royal. Soc, A(1966),
289, 301-320.
\bibitem{fetter} A.L. Fetter, Phys.Rev. A {\bf 53} 4246 (1996).
\bibitem{NazarenkoNewell} L. Biven, S.V. Nazarenko and A.C. Newell, ``
Breakdown of wave turbulence and the onset of intermittency'', Physics
Letters A, vol.280, no.1-2 pp. 28-32
\bibitem{Z68a} V.E.Zakharov, Zh. Priklad. Tech. Fiz, 1, 35 (1965).
J.Appl.Mech. Tech. Phys, {\bf 1}, 22, (1967)
\bibitem{Z68b} V.E.Zakharov Zh.Eksp.Teor.Fiz. {\bf 51},688,(1966)
Sov.Phys.JETP {\bf 24} (1967).
\bibitem{Bogolubov1947} Bogoliubov N, J. Phys (Moscow), {\bf 11}, 23,
(1947).
\bibitem{Newell68} A.C. Newell: Rev. Geophys. {\bf{6}}, 1 (1968)
\bibitem{nkd} S. Nazarenko, N. Kevlahan and B. Dubrulle, Journal of
Fluid Mech. {\bf 390}, 325, (1999).
\bibitem{LBN} Yuri Lvov, R. Binder and Alan Newell, Physica D, {\bf 121}, pp. 317 - 343,(1998).
\bibitem{GLBDZ98} C.W. Gardiner, M.D. Lee, R.J. Ballagh, M.J. Davis,
P. Zoller: PRL \textbf{81}, 24 (1998).
\bibitem{naz-lvov} S. Nazarenko and Y. Lvov, Normal Hamiltonian forms
for linear waves in weakly inhomogeneous media, to be submitted to
Journal of Math Phys.
\bibitem{landafshits10} E.M. Lifshits and L.P. Pitaevskii, Physical
Kinetics
\bibitem{bbk} B.B. Kadomtsev, Plasma turbulence (Academic Press, New
York, 1965).
\bibitem{tsit1} V.N. Tsytovich, Nonlinear effects in plasma (Nauka,
Moscow, 1967).
\bibitem{tsit2} V.N. Tsytovich, Theory of a turbulent plasma
(Consultants Bureau, New York, 1977)
\bibitem{zmrub} V.E. Zakharov, S.L. Musher and A.M. Rubenchik, Phys
Reports, {\bf 129}, 285 (1985).
\bibitem{papaLvov} V.S. Lvov, Nonlinear Spin Waves, p. 192 (Nauka,
Moscow, 1987).
\bibitem{Newell_ann} A.C. Newell, S.V. Nazarenko and L. Biven, Physica
D, {\bf 152-153} 520-550 (2001) 
\end{thebibliography}
 \end{document}